\documentclass[12pt]{iopart}
\usepackage{graphicx}
\usepackage{caption}
\usepackage{subcaption}
\usepackage{color}
\usepackage{amsbsy}
\usepackage{color}
\usepackage{epsfig}
\usepackage{graphicx}
\usepackage{amssymb}
\usepackage{bbm}
\usepackage{amsbsy}
\usepackage{slashed}
\usepackage{iopams}
\usepackage{xcolor}

\newcommand{\beq}{\begin{equation}}
\newcommand{\eeq}{\end{equation}}
\definecolor{mbcol}{rgb}{1,0,1}

\usepackage[normalem]{ulem}

\begin{document}
\title[]{Schwinger Pair Production in QCD from Flavor-Dependent  Contact Interaction Model of Quarks}
\author{Aftab Ahmad,~Akif Farooq}
\address{Institute of Physics, Gomal University, 29220, D.I. Khan, Khyber Pakhtunkhaw, Pakistan.}
\ead{aftabahmad@gu.edu.pk, akiffarooq143@gmail.com}

\begin{abstract}
We study the Schwinger mechanism in QCD i.e., the quark-antiquark pair production rate $\Gamma$  in the presence of pure electric field strength $eE$, for a higher number of colors $N_c$ and flavors $N_f$. In this context, our unified formalism is based on the Schwinger-Dyson equations,  flavor-dependent symmetry preserving
vector-vector contact interaction model of quarks, and an optimal time regularization scheme. For fixed $N_c=3$ and $N_f=2$, the dynamically quark mass decreases as we increase $eE$ and near at and above the  pseudo-critical electric field $eE_c$, the chiral symmetry is restored and quarks becomes unconfined. The pair production rate $\Gamma$ becomes stable  and grows quickly  above $eE_c$.  For fixed $N_c=3$ and upon increasing $N_f$ the dynamical mass suppresses and as a result, the $eE_c$ reduces to its smaller values, the pair production rate $\Gamma$ tends to initiates and  grows quickly for smaller values of $eE_c$. In contrast, for fixed $N_f=2$ and upon increasing $N_c$, the dynamical chiral symmetry is restored for larger and larger values of $eE_c$ and at $N_c\geq4$, the transition changes from smooth cross-over to the first order at some critical endpoint ($N_{c,p}, eE_{c,p}$). Consequently, the quark-antiquark production rate $\Gamma$ needs higher values of  $eE_c$ for the stable and  quick growth as we increase $N_c$. 
 Our findings are satisfactory and in agreement with already predicted results for pair production rate (for fixed $N_c=3$ and $N_f=2$) by other reliable effective models of QCD.
\end{abstract}


\textit {}

\maketitle

\section{Introduction} \label{section-1}
Historically, the phenomenon of electron-positron pair production from a non-perturbative vacuum in the presence of electric field strength $eE$ (where $e$ is an electric charge)  was first introduced by Fritz Sauter~\cite{Sauter:1931zz}, Heisenberg and Euler~\cite{Heisenberg:1936nmg}. However, the complete theoretical framework provided by Julian Schwinger in the field of quantum electrodynamics (QED)~\cite{Schwinger:1951nm}, and named after him as the Schwinger effect or Schwinger mechanism. Later on, this phenomenon has been widely studied in the field of quantum chromodynamics (QCD), see for example, Refs.~\cite{Yildiz:1979vv, Cox:1985bu, Suganuma:1990nn,Suganuma:1991ha,Tanji:2008ku}. As we know that the QCD  is a theory of strong color force among the quarks and gluons possess two special properties: The asymptotic freedom (i.e., the quarks interacts weakly
at a short distancesor high energy scale) and the quark confinement  
(i.e., the quarks interacts strongly  at a large distance or low energy scale).  
The Schwinger effect in QCD is thus, the production of quark-antiquark pairs in the presence of a strong electric field $eE$. The  quark-antiquark pair production can be calculated from  the Schwinger pair production rate $\Gamma$ and is, defined as the probability, per unit time and per unit
volume that a quark-antiquark pair is created by the constant electric
field $eE$. \\Another important property of low-energy QCD is 
the dynamical chiral symmetry breaking,  which is related to the dynamical mass generation of the quarks. The strong electric field tends to restore the dynamical chiral symmetry and as a result,  the dynamically generated quark mass suppresses with the increase of $eE$.  It can be understood by realizing that being closer together, the quark and antiquark pairs are reaching the asymptotic freedom regime faster by reducing the interaction strength as the intensity of the electric field $eE$ increases. Such a phenomenon is sometimes referred to as the chiral electric inhibition effect~\cite{Klevansky:1988yw, Suganuma:1990nn, Klimenko:1992ch, Klevansky:1992qe, Babansky:1997zh, Cao:2015dya, Tavares:2018poq}, or the chiral electric rotation effect~\cite{Wang:2017pje} or {inverse electric catalysis}(IEC)~\cite{Ruggieri:2016lrn, Tavares:2019mvq, Ahmad:2020ifp}. The nature of the dynamical chiral symmetry breaking--restoration and confinement--deconfinement phase transition is of second-order when the bare quark mass $m=0$ (i.e., in the chiral limit) while cross-over when  $m\neq0$.  It has been argued in Refs. ~\cite{Cao:2015dya, Tavares:2018poq, Tavares:2019mvq}  that the pair production rate $\Gamma$, increases quickly  near some pseudo-critical electric field $eE_c$, where the chiral symmetry is restored. \\  
The study of the effect of the electric field on the chiral phase transition plays a significant role in Heavy-Ion Collision experiments. In such experiments, the magnitude of the electric and magnetic fields produced with the same order of magnitude ($\sim 10^{18}$ to $10^{20}$Gauss)~\cite{Bzdak:2011yy, Deng:2012pc, Bloczynski:2012en, Bloczynski:2013mca} in the event-by-event collisions using Au $+$ Au at  RHIC-BNL, and in a non-central Heavy-Ion collision of  Pb $+$ Pb in ALICE-LHC. Besides, experiments with asymmetric
Cu $+$ Au collisions, it is believed that the strong
electric field is supposed to be created in the overlapping region~\cite{Hirono:2012rt, Voronyuk:2014rna, Deng:2014uja}. It happens because there are different numbers
of electric charges in each nuclei, which may be
due to the charge dipole formed in
the early stage of the collision. Some other phenomena like the chiral electric separation effect~\cite{Huang:2013iia, Jiang:2014ura}, the particle polarization effect~\cite{Karpenko:2016jyx, Xia:2018tes, Wei:2018zfb}, etc., which may emerge due to the generation of vector and/or axial current in the presence of strong electromagnetic fields. \\ It is illustrative to approximate the created number of
charged quark-antiquark pairs in the 
QGP produces in Heavy Ion collision, because in the QGP phase the dynamical chirl symmetry restores and the deconfinement occurs.
According to advance numerical
simulations, the electric fields created in Au $+$ Au
collisions at center-of-mass energy $\sqrt{s}=200$ GeV 
 due to the fluctuation is of
the order $eE \sim m^{2}_{\pi}$
while in Pb$+$Pb collisions at
$\sqrt{s}=2.76$ TeV is of the order $eE\sim20 m^{2}_{\pi}$
\cite{Deng:2012pc}. Then by assuming
the space-time volume of the QGP is of the order $\sim (5 {\rm fm})^4$,
the total pair creation number is $N_{\rm RHIC}=3.5$, and $N_{\rm LHC}=1400$ \cite{Cao:2015xja}, gives us a clear indication of the importance of Schwinger pair production of quark
and antiquark in Heavy Ion Collisions.\\ 
It is well understood that the QCD exhibits confinement and chiral symmetry breaking with 
the small number of light quark 
flavors $N_f$. However, for larger
$N_f$, Lattice QCD simulation~\cite{LSD:2014nmn,Hayakawa:2010yn,Cheng:2013eu,Hasenfratz:2016dou,LatticeStrongDynamics:2018hun},
as well as the continuum methods of QCD ~\cite{bashir2013qcd,Appelquist:1999hr,Hopfer:2014zna,Doff:2016jzk,Binosi:2016xxu,Ahmad:2020jzn,Ahmad:2020ifp,Ahmad:2022hbu}, predicted that there is a critical value
$N^{c}_{f}\approx8$
above which the chiral symmetry is restored and quarks become
unconfined.  It has been discussed in detail in  Ref.~\cite{Ahmad:2020ifp} that the critical number of flavors $N^{c}_{f}$ suppresses with the increase of temperature $T$ and enhances with the increasing magnetic field $eB$.  Even the QCD phase diagram at finite temperature $T$ and density $\mu$  suppresses with the increase of light quark flavors, see for an instance Ref.~\cite{Ahmad:2020jzn}. Besides the higher number of light quark flavors, QCD with a larger number of colors $N_c$ in the
fundamental $SU(N_c)$ representation also plays a  significant role.  It has been demonstrated in Ref.~\cite{Ahmad:2020ifp, Ahmad:2020jzn} that the 
chiral symmetry is dynamically broken above a critical value $N^{c}_{c}\approx2.2$, as a result, the dynamically generated mass increases 
near and above $N^{c}_{c}$. Increasing the number of colors also enhances the critical temperature $T_c$ and the critical chemical potential $\mu_c$ of the chiral phase transition in the QCD phase diagram ~\cite{Ahmad:2020ifp}. Both $N_c$ and magnetic field $eB$ strengthen the generation of  the dynamical masses of the quarks
~\cite{Ahmad:2020jzn}.\\  
It will be more significant to study the dynamical chiral symmetry breaking and Schwinger effect in pure electric field background for a higher number of light quark flavors $N_f$  and for a large number of colors $N_c$, which has not yet been studied so far, as far as we know. It may have a stronger impact not only on the theoretical ground but also in Heavy-Ion Collision experiments where a large number of light flavors of quark-antiquark pairs are produced.
Our main objective of this work is to study the quark-antiquark pair production rate in the presence of a pure electric field $eE$ background for a higher number of light quark flavors $N_f$ and colors $N_f$. For this purpose, we use the  Schwinger-Dyson equation in the rainbow-ladder truncation, in the Landau gauge, the symmetry preserving flavour-dependent confining vector-vector contact interaction model of quarks \cite{Ahmad:2020ifp},
and the Schwinger optimal time regularization scheme \cite{Schwinger:1951nm}. The pseudo-critical electrical field strength $eE_c$, the critical number of flavors $N^{c}_{f}$ and the critical number of colors $N^{c}_{c}$ for chiral symmetry breaking-restoration can be obtained from the peak of the correspondence gradient of dynamical quark mass, whereas the confinement-deconfinement can be triggered from the peaks of correspondence gradient of the confining length scale~\cite{Ahmad:2016iez, Ahmad:2020ifp, Ahmad:2020jzn}. It
should be noted that the chiral symmetry restoration and
deconfinement occur simultaneously in this model~\cite{Marquez:2015bca, Ahmad:2016iez, Ahmad:2020ifp, Ahmad:2020jzn}. \\ 
This manuscript is organized as follows: In Sec.~2, we present the general formalism for the flavor-dependent contact interaction model and QCD gap equation.  In Sec.~3, We discuss the gap equation and the the Schwinger pair production rate in the presence of electric field $eE$ for a large number of flavors $N_f$ and colors $N_c$. In Sec.~4,  we present the numerical solution of the gap equation and the Schwinger pair production rate in the presence of $eE$ for a higher number of $N_f$ and $N_c$. In the last Sec.~5, we present the summary and future perspective of our work.
\section{General formalism and flavor-dependent Contact Interaction model} \label{section-2}
The Schwinger-Dyson equations (SDE)  for  the dressed-quark propagator $S$,  is given by:
\begin{eqnarray}
S^{-1}(p)&=S^{-1}_{0}(p) + \Sigma(p)\,,\label{CI1}
\end{eqnarray}
where $S_{0}(p)=(\slashed{p}+ m + i\epsilon)^{-1}$, is the bare quark propagator and $S(p)=(\slashed{p}+ M + i\epsilon)^{-1}$ is the dressed quark propagator.   The $\Sigma(p)$ is the self energy and is given by
\begin{eqnarray}
\Sigma(p)=\int \frac{d^4k}{(2\pi)^4} g^{2}
 D_{\mu\nu}(q)\frac{\lambda^a}{2}\gamma_\mu S(k)
\frac{\lambda^a}{2}\Gamma_\nu(p,k)\,,\label{CI2}
\end{eqnarray}
where $\Gamma_\nu (k,p)$  is the dressed quark-gluon vertex and $g^{2}$ is the QCD coupling constant. The
$D_{\mu\nu}(q)= D(q)(\delta_{\mu\nu} -\frac{q_{\mu} q_{\nu}}{q^2})$ is the gluon propagator in the Landau gauge with $\delta_{\mu \nu}$ is the metric tensor in  Euclidean space, $D(q)$ is the gluon scalar function and  $q=k-p$ is the gluon four momentum. 
The $m$ is the current quark mass,  which may set equal to zero (i.e., $m=0$) in the chiral limit.  The $\lambda^a$'s are the
Gell-Mann matrices and in the ${\rm SU(N_c)}$ representation, the Gell-Mann's  matrices satisfies the following identity:
\begin{eqnarray}
\sum^{8}_{a=1}\frac{{\lambda}^a}{2}\frac{{\lambda}^a}{2}=\frac{1}{2}\left(N_c - \frac{1}{N_c} \right)I, \label{CI3}
\end{eqnarray} 
here $I$ is the unit matrix. In this work, we use   the  symmetry preserving  flavor-dependent  confining contact interaction  model~\cite{Ahmad:2020jzn, Ahmad:2022hbu} for the gluon propagator (in Landau gauge) in the infrared region where the gluons dynamically acquire a mass $m_{g}$~\cite{Langfeld:1996rn, Cornwall:1981zr,Aguilar:2015bud,GutierrezGuerrero:2010md, Kohyama:2016obc}, is given by
\begin{eqnarray}
  g^2 D_{\mu \nu}(k) &=&  \frac{4 \pi
  \alpha_{\rm ir}}{m_G^2}\sqrt{1 - \frac{(N_{f}-2)}{\mathcal{N}_{f}^{c}}} \delta_{\mu \nu} =\delta_{\mu \nu} \alpha_{\rm eff} \sqrt{1 - \frac{(N_{f}-2)}{\mathcal{N}_{f}^{c}}}\,,\label{CI4}
\end{eqnarray}
the $\alpha_{\rm
ir}=0.93\pi $ is the infrared enhanced interaction strength parameter, $m_g=800$ MeV is the gluon mass scale~\cite{Boucaud:2011ug}.   The $\mathcal{N}_{f}^{c}=N^c_{f}+\eta$ is some guess values of critical number of 
flavors. In Ref.~\cite{Ahmad:2020jzn, Ahmad:2022hbu}, the value of $\eta$   has been set and it ranging from $1.8-2.3$, to obtained the desired number of critical number $N^{c}_{f}\approx8$ above which the dynamical symmetry restored and deconfinement occurred. It has been
argued in the Ref.~\cite{Ahmad:2020jzn} that the appearance of the parameter $\eta$ is
because of the factor $(N_f-2)$ in Eq.~(\ref{CI4}).\\
With a particular choice of the flavor-dependent model Eq.~(\ref{CI4}),  the dynamical quark mass function is merely a constant and  the
the dressed quark propagator takes 
into a very simple form~\cite{Ahmad:2015cgh}:
\begin{eqnarray}
S^{-1}(p)=i\gamma\cdot p+M_f\;.\label{CI5}
\end{eqnarray}
It is because the wave function renormalization trivially
tends to unity in this case, and the quark mass function $M$
become momentum independent:
\begin{equation}
M_f=m_f+\frac{\alpha_{\rm eff} \alpha^{N_c}_{N_f}}{2}\int^\Lambda \frac{d^4k}{(2\pi)^4} {\rm Tr}[S_f(k)]\;.\label{CI6}
\end{equation}
Where $M_f$ is the dynamical mass and  $\alpha^{N_c}_{N_f}=\sqrt{1 - \frac{(N_{f}-2)}{\mathcal{N}_{f}^{c}}}\left(N_c -\frac{1}{N_c}\right)$. After simplifying Eq.~(\ref{CI6}), we have 
\begin{eqnarray}
M_f=m_f+2\alpha_{\rm eff} \alpha^{N_c}_{N_f}\int \frac{d^4k}{(2\pi)^4} \frac{M}{k^2+M^2}\;.\label{CI7}
\end{eqnarray}
The  quark-anitquark condensate which serves as an order parameter for the dynamical chiral symmetry breaking in this truncation,  can be written as
\begin{eqnarray}
-\langle \bar{q} q\rangle= \frac{M_f-m_f}{2\alpha_{\rm eff}}\;.\label{CI8}
\end{eqnarray}
Using $d^4 k= (1/2) k^2 dk^2 \sin ^2 \theta  d\theta \sin \phi d \phi d\psi $, performing the trivial integration's and using the variable $s=k^2$ in  Eq.~(\ref{CI7}), we have
\begin{eqnarray}
M_f=m_{f}+\frac{\alpha_{\rm eff}\alpha^{N_c}_{N_f}}{8\pi^2}\int^{\infty }_{0}
ds\frac{s}{s+M_f^2} \,. \label{CI9}
\end{eqnarray}
The above integral in Eq.~(\ref{CI9}) is divergent and  we need to regularize it.  In the present scenario we use the Schwinger proper-time regularization procedure~\cite{Schwinger:1951nm}. In this procedure, we take the exponent 
integrand's denominator and then introduce an additional
infrared cutoff, in addition to the conventional ultraviolet that is normally used in NJL model Studies.
Accordingly,  the   confinement is  implemented through an infrared cut-off
~\cite{Ebert:1996vx}. The significance of adopting the mentioned regularization procedure is,
the quadratic and logarithmic divergences remove and the axial-vector Ward-Takahashi identity~\cite{Ward:1950xp, Takahashi:1957xn} is satisfied.
From  Eq.~(\ref{CI9}),
the integrand's denominator reduced to:
\begin{eqnarray}
\frac{1}{s+M^{2}_{f}}&=&\int^{\infty }_{0} d\tau {\rm e}^{-\tau(s+M^{2}_{f})}
\rightarrow
 \int^{\tau_{ir}^2}_{\tau_{uv}^2} d\tau
 {\rm e}^{-\tau(s+M^{2}_{f})} \nonumber\\ &=&
\frac{ {\rm e}^{-\tau_{uv}^2(s+M^{2}_{f})}-{\rm e}^{-\tau_{ir}^2(s+M^{2}_{f})}}{s+M^{2}_{f}}\;. \label{CI10}
\end{eqnarray}
Here, $\tau_{uv}=\Lambda^{-1}_{uv}$ is an ultra-violet regulator, which plays the dynamical role and sets the scale for all dimensional quantities.
The $\tau_{ir}=\Lambda^{-1}_{ir}$ stands for the infrared regulator whose non zero value implements confinement by ensuring the absence of quarks production thresholds ~\cite{Roberts:2007jh}. Hence  $\tau_{ir}$ referred to as the confinement scale~\cite{Ahmad:2016iez}. From Eq.~(\ref{CI10}), it is now clear that the location of the original pole is at $s=-M^2$, which is canceled by the numerator.
Thus the propagator is free from real as well as the complex poles, which is consistent with the definition of confinement i.e., ``an excitation
described by a pole-less propagator would never reach its
mass-shell''~\cite{Ebert:1996vx}. \\
After integration over `s', the gap equation Eq.~(\ref{CI9}) is reduced to:
\begin{eqnarray}
 M_f&=& m_{f} + \frac{M_f^3 \alpha_{\rm eff}\alpha^{N_c}_{N_f}}{8\pi^{2}}
  \Gamma(-1,\tau_{uv} M_{f}^2,\tau_{ir} M_{f}^2)\,,\label{CI11}
\end{eqnarray}
\noindent where
\begin{equation}
\Gamma (a, y_1,y_2)=\Gamma (a,y_1)-\Gamma(a,y_2)\,\label{CI12}
\end{equation}
with $\Gamma(a,y) = \int_{y}^{\infty} t^{\alpha-1} {\rm
e}^{-t} dt$, is the incomplete Gamma function. The above equation Eq.~(\ref{CI11}), is the gap equation in vacuum which is regularized in the Schwinger proper time regularization scheme with two regulators.
The confinement in this model can be triggered from the confining length scale~\cite{Ahmad:2016iez, Ahmad:2020ifp, Ahmad:2020jzn}:
\begin{eqnarray}
\tilde{\tau}_{ir}=\tau_{ir}\frac{M}{M_f},\label{CI13}
\end{eqnarray}
where $M=M(3,2)$ is the dynamical mass for fixed $N_c=3$,  $N_c=2$. Here ${M}_f=M(N_c,N_f)$  is the generalized color $N_c$ and flavor $N_f$ dependent dynamical mass.  As, the~$\tau_{ir}$ is introduced in the
model to mimic confinement by ensuring the absence of quarks production thresholds, so in the  presence of $N_f$ and  $N_c$, it 
is required to vary slightly  with both $N_f$ and $N_c$.  Thus the entanglement between dynamical chiral symmetry breaking and confinement is expressed through an explicit $N_f$ and  $N_c$-dependent regulator i.e., $\tilde{\tau}_{ir}=\tau_{ir}(N_c,N_f)$, in the infrared.
For chiral quarks (i.e., $m_f=0$), the confining scale $\tilde{\tau}_{ir}$ diverges at the chiral symmetry restoration region.
Next we discuss the general formalism  of the gap equation and the Schwinger pair production rate in the  in the presence of electric field and in the presence of electric field  $eE$ and for higher $N_f$ and $N_c$.\\
\section{Gap equation and Schwinger pair production rate in the presence of electric field}
In this section, we discuss the gap equation in the presence of a uniform and homogeneous pure electric field $eE$. In QCD Lagrangian, the interaction  with  pure electric field $A^{ext}_\mu$  embedded in the  covariant derivative,
\begin{eqnarray}
 D_\mu=\partial_\mu -iQ_f A_\mu^{\rm ext}, \label{em1}
\end{eqnarray}
with $Q_{f}=(q_u=+2/3 ,q_d=-1/3)e$ is refers to the electric charges of $u$ and $d$-quark, respectively. We choose  the symmetric gauge  vector potential $A^{ext}_\mu=-\delta_{\mu0}x_{3} E
$, to obtain the resulting electric field along the z-axis. The gap equation in the presence of pure electric field continues to form Eq.~(\ref{CI6}), where $S_f(k)$ is now  dressed
with electric field $eE$, that is, $ {S_f}(k) \rightarrow\tilde{S_f}(k)$~\cite{Schwinger:1951nm, Klevansky:1992qe, Cao:2015dya, Wang:2017pje, Cao:2021rwx}, and is given as
\begin{eqnarray}
&&\hspace{-22mm}\tilde{S_f}(k)=\int^{\infty}_{0} d\tau {\rm e}^{-\tau \bigg( M_{f}^{2}+(k_{4}^{2}+k_{3}^{2}) \frac{{\rm tan}(|Q_{f}E| \tau)}{|Q_{f}E|\tau}+k_{1}^{2}+k_{2}^{2}\bigg)}
\nonumber\\&&\hspace{-20mm}\times \bigg[ M_f+i\gamma^{4}\bigg(k_4 +{\rm tan}(|Q_{f}E\tau |)\bigg)k_3 -\gamma^{3}(k_3 +{\rm tan}(|Q_{f}E\tau |)k_4)-\bigg(\gamma^{2}k_2-\gamma^{1}k_1 \bigg)
\nonumber\\&&\hspace{-20mm}\times
\bigg(1+{\rm tan}(|Q_{f}E|\tau)\gamma^4 \gamma^3 \bigg) \bigg]\,.\label{em2}
\end{eqnarray}
Where $\gamma$'s are the Dirac gamma matrices. Taking the trace ``Tr'' of the of the propagator Eq.~(\ref{em2}), inserting it in  Eq.~(\ref{CI6}) and after carrying out the  the integration over $k$'s, 
the gap equation in the electric field $eE$ and  with the flavor-dependent contact interaction model of quark~\cite{Ahmad:2020jzn} is given by
\begin{eqnarray}
M_f &=& m_{f}+ \frac{\alpha_{\rm eff} \alpha^{N_c}_{N_f}}{8\pi^2}\sum_{f} \int^{\infty}_{0} \frac{d\tau}{\tau^2} M_f {\rm e}^{-\tau M_{f}^{2}}
\bigg[\frac{|Q_{f}E|}{{\rm tan}(|Q_{f}E|\tau)} \bigg]\,.\label{em3}
\end{eqnarray}
Next, we separate the vacuum and  electric field dependent part by using the vacuum  subtraction scheme\cite{Klevansky:1992qe, Cao:2015dya}, as given by    
\begin{eqnarray}
M_f&=& m_f+\frac{\alpha_{\rm eff} \alpha^{N_c}_{N_f}}{8\pi^2}\sum_{f} \int^{\infty}_{0} \frac{d\tau}{\tau^2} M_f {\rm e}^{-\tau M_{f}^{2}}\nonumber\\&&\hspace{-0.5mm}+ \frac{\alpha_{\rm eff} \alpha^{N_c}_{N_f}}{8\pi^{2}}\sum_{f} \int^{\infty}_{0} \frac{d\tau}{\tau^2} M_f {\rm e}^{-\tau M_{f}^{2}}
\bigg[\frac{|Q_{f}E|\tau}{{\rm tan}(|Q_{f}E|\tau)}-1\bigg]\,.\label{em4}
\end{eqnarray}
The vacuum integral can be regularized with the two regulators as in Eq.~(\ref{CI11}), and thus the  Eq.~(\ref{em4}) can be reduced to: 
\begin{eqnarray}
M_f&=& m_f+\frac{M_{f}^{3} \alpha_{\rm eff} \alpha^{N_c}_{N_f}}{8\pi^{2}}
\Gamma(-1,\tau^{2}_{uv} M_{f}^2,\tau^{2}_{ir} M_{f}^2)\nonumber\\&&\hspace{-0.5mm}+ \frac{\alpha_{\rm eff} \alpha^{N_c}_{N_f}}{8\pi^{2}}\sum_{f} \int^{\infty}_{0} \frac{d\tau}{\tau^2} M_f {\rm e}^{-\tau M_{f}^{2}}
\bigg[\frac{|Q_{f}E|\tau}{{\rm tan}(|Q_{f}E|\tau)}-1\bigg]\,.\label{em5}
\end{eqnarray}
The Eq.~(\ref{em5}) can also be written as 
\begin{eqnarray}
\frac{M_f-m_f}{\alpha_{\rm eff}}&=&\frac{M_{f}^{3}  \alpha^{N_c}_{N_f}}{8\pi^{2}}
\Gamma(-1,\tau^{2}_{uv} M_{f}^2,\tau^{2}_{ir} M_{f}^2)\nonumber\\&&\hspace{-0.5mm}+ \frac{ \alpha^{N_c}_{N_f}}{8\pi^{2}}\sum_{f} \int^{\infty}_{0} \frac{d\tau}{\tau^2} M_f {\rm e}^{-\tau M_{f}^{2}}
\bigg[\frac{|Q_{f}E|\tau}{{\rm tan}(|Q_{f}E|\tau)}-1\bigg]\,.\label{em6a}
\end{eqnarray}
As we have  already  regularized the integral in Eq.~(\ref{em5}), using the vacuum subtraction scheme but we still have poles associated with the tangent term in the denominator of  our gap
equation when $|Q_fE|\tau=n\pi$ with $n=1,2,3.....,$. Upon taking the principle value (real value) of the integral~\cite{Klevansky:1992qe},
given in Eq.~(\ref{em6a}), we have
\begin{eqnarray}
\int^{\infty}_{0} \frac{d\tau} {\tau^2}{\rm e}^{-\tau M_{f}^{2}}
\bigg(\frac{|Q_{f}E|\tau}{{\rm tan}(|Q_{f}E|\tau)}-1\bigg)&=& |Q_{f}E| Re J(i M^{2}_{f}/2|Q_{f}E|)\,,\label{em6b}
\end{eqnarray}
with
\begin{eqnarray}
J(\zeta) = 2i  \bigg[(\zeta -\frac{1}{2}){\rm log}\zeta-\zeta-{\rm log}\Gamma(\zeta)+\frac{1}{2}{\rm log}2\pi] \bigg]\,.\label{em6c}
\end{eqnarray}
The effective potential $\Omega$, can be obtained by integrating the Eq.~(\ref{em6a}) over dynamical mass $M_f$:
\begin{eqnarray}
\Omega &=&\frac{(M_f-m_f)^2}{2\alpha_{\rm eff}}-\frac{\alpha^{N_c}_{N_f}{M_f}^{4}}{32\pi^{2}}
  \bigg[\frac{{\rm e}^{-\tau M_{f}^{2}}}{\tau^{4}_{uv}}-\frac{{\rm e}^{-\tau M_{f}^{2}}}{\tau^{4}_{ir}}-M_{f}^4\Gamma(-1,\tau^{2}_{uv} M_{f}^2,\tau^{2}_{ir} M_{f}^2)\bigg]\nonumber\\&&\hspace{-0.5mm}-\frac{ \alpha^{N_c}_{N_f}}{16\pi^{2}}\sum_{f} \int^{\infty}_{0} \frac{d\tau}{\tau^3} M_f {\rm e}^{-\tau M_{f}^{2}}
\bigg[\frac{|Q_{f}E|\tau}{{\rm tan}(|Q_{f}E|\tau)}-1\bigg]\,.\label{em7a}
\end{eqnarray}
We noted that there 
are infinite poles in the  effective potential $\Omega$ too which are due to the contribution of the Schwinger pair
production in the presence of the tangent (electric field-related) term    
. These poles yields the effective
potential to be complex with its  imaginary part giving the
Schwinger pair production rate~\cite{Schwinger:1951nm,Ruggieri:2016lrn, Cao:2015dya, Tavares:2018poq}.
The third term in the Eq.~(\ref{em7a}) can be further simplified by using the following trigonometric relation:
\begin{eqnarray}
\frac{\pi\tau}{{\rm tan}(\pi\tau)}-1 =\sum_{n=1}^{\infty} \frac{2 \tau^2}{\tau^2-n^2}\,.\label{em7b}
\end{eqnarray}
Using Eq.~(\ref{em7b}) in Eq.~(\ref{em7b}), we have
\begin{eqnarray}
\Omega &=&\frac{(M_f-m_f)^2}{2\alpha_{\rm eff}}-\frac{\alpha^{N_c}_{N_f}{M_f}^{4}}{32\pi^{2}}
  \bigg[\frac{{\rm e}^{-\tau M_{f}^{2}}}{\tau^{4}_{uv}}-\frac{{\rm e}^{-\tau M_{f}^{2}}}{\tau^{4}_{ir}}-M_{f}^4\Gamma(-1,\tau^{2}_{uv} M_{f}^2,\tau^{2}_{ir} M_{f}^2)\bigg]\nonumber\\&&\hspace{-0.5mm}-\frac{ \alpha^{N_c}_{N_f}}{16\pi^{2}}\sum_{f} \int^{\infty}_{0} \frac{d\tau}{\tau^3} M_f {\rm e}^{-\tau M_{f}^{2}}
\sum_{n=1}^{\infty} \frac{2\tau^2}{\tau^2-\frac{\pi^{2}n^2}{(|Q_{f}E|)^2}}\,.\label{em8}
\end{eqnarray}
The  Schwinger pair production rate $\Gamma$,  corresponds to the imaginary
part of the effective potential Eq.~(\ref{em8}), and is thus, given by
\begin{eqnarray}
\Gamma=-2{\rm Im}\Omega=\sum_{f}\sum^{\infty}_{n=1}\frac{\rm \alpha^{N_c}_{N_f}} {4\pi}\frac{(|Q_f E|)^{2}{\rm e}^{-n\pi M^{2}_f/|Q_f E|}}{(n\pi)^2}\,,\label{em9}
\end{eqnarray}
which does not depend on the number of color $N_c$ or the number of flavors $N_f$ explicitly. However, since the dynamical quark mass $M_f$ depends
on both $N_f$  and $N_c$  as it is obvious from the gap equation Eq.~(\ref{em4}), they will
affect the quark-antiquark pair production rate $\Gamma$ implicitly through $M_f$. The numerical solution of the gap equation and the Schwinger pair production rate  will be discussed in the next section.
\section{Numerical Results} 
\begin{figure}
\begin{subfigure}[b]{7cm}
\centering\includegraphics[width=8cm]{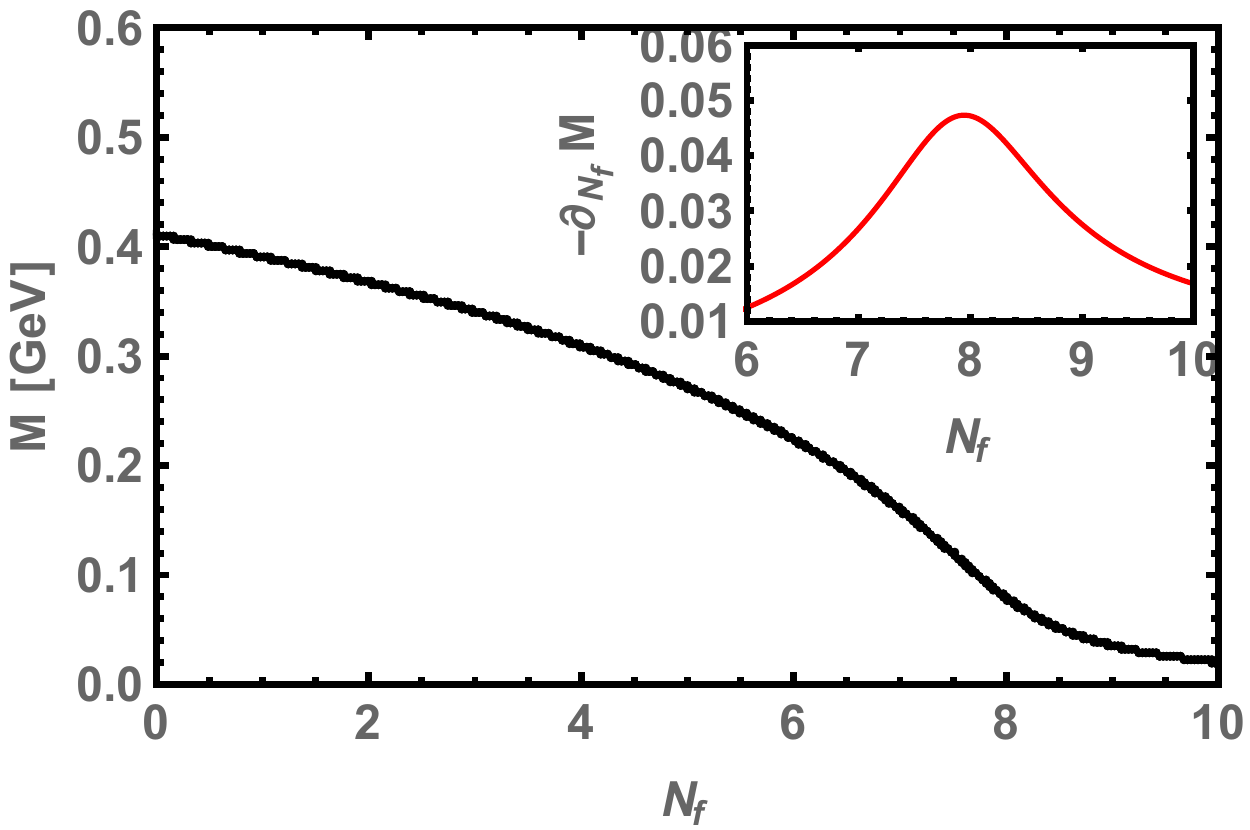}
\caption{}\label{fig:a}
\end{subfigure}
\hfill
\begin{subfigure}[b]{7cm}
\centering\includegraphics[width=8cm]{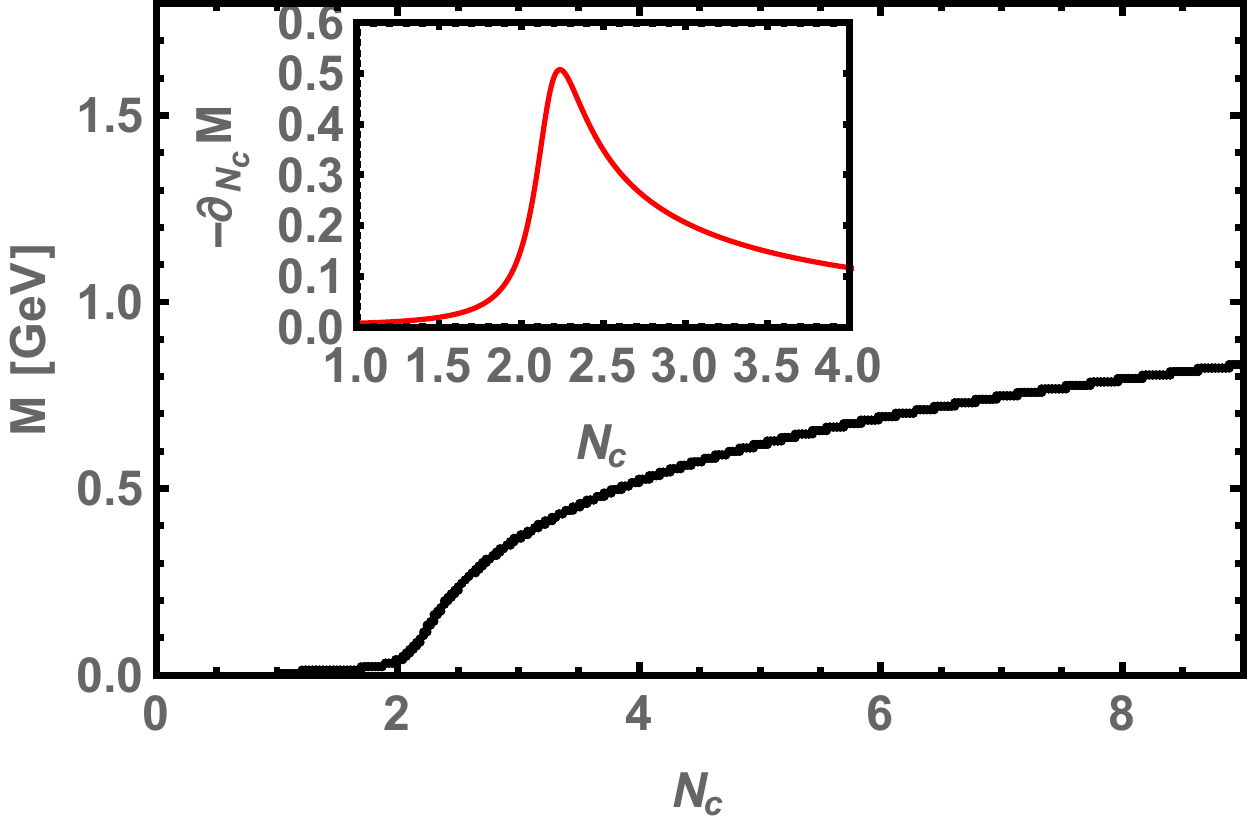}
\caption{}\label{fig:b}
\end{subfigure}
\caption{(\subref{fig:a}) Behavior of the dynamical  quark mass $M$  and its flavor-gradient $\partial_{N_f}M$ as a function of number of flavors $N_f$  for fixed  number of  $N_c=3$.   The dynamical mass $M$ decreases as we increase $N_f$ and at some critical value $N^{c}_f=8$, the dynamical chiral symmetry is restored. \\ (\subref{fig:b})  The  behavior of the dynamical  quark mass $M$  and its color-gradient  $\partial_{N_c}M$ as a function of number of colors $N_c$ and   for fixed $N_f=2$.   From the peak of the  $\partial_{N_c}M$, it is clear that the  dynamical chiral symmetry is broken above  $N^{c}_c=2.2$.}
  \label{fig1}
\end{figure}

\begin{figure}
\begin{subfigure}[b]{7cm}
\centering\includegraphics[width=8cm]{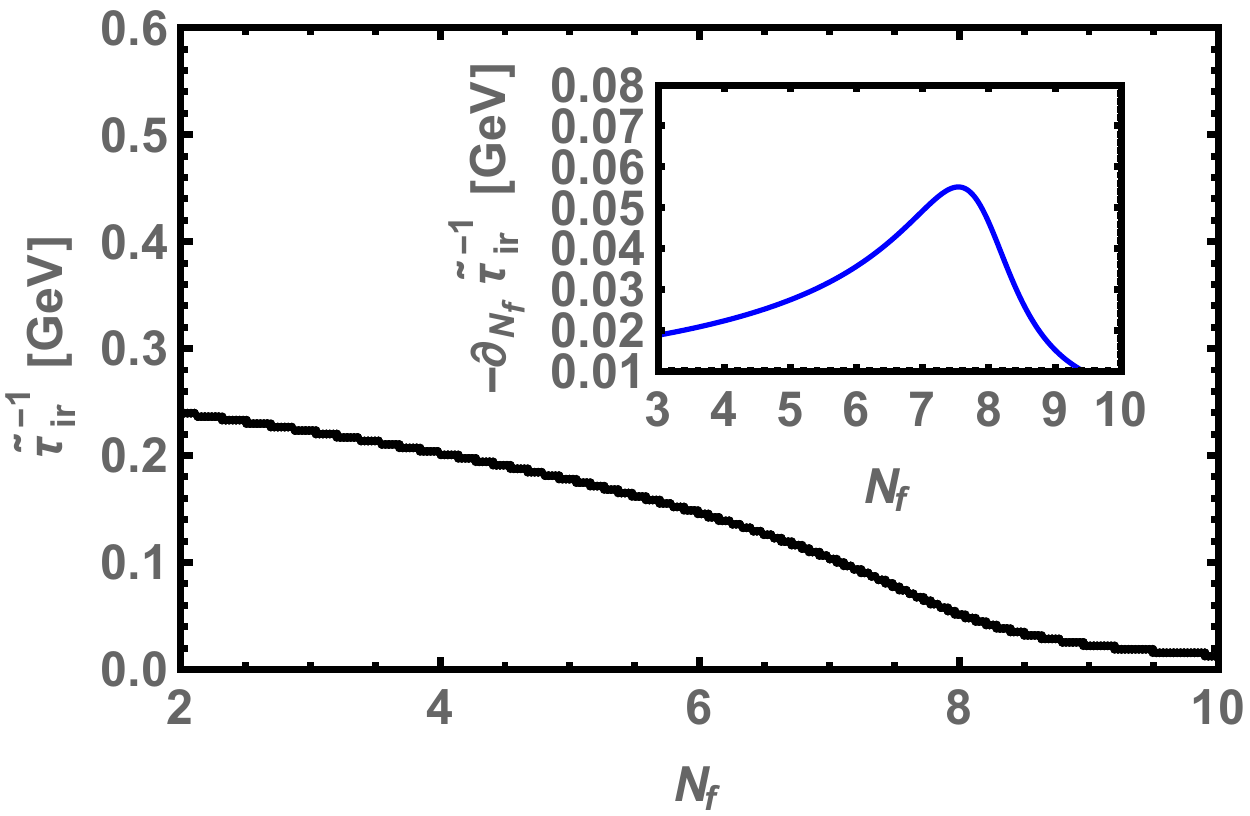}
\caption{}\label{fig:a}
\end{subfigure}
\hfill
\begin{subfigure}[b]{7cm}
\centering\includegraphics[width=8cm]{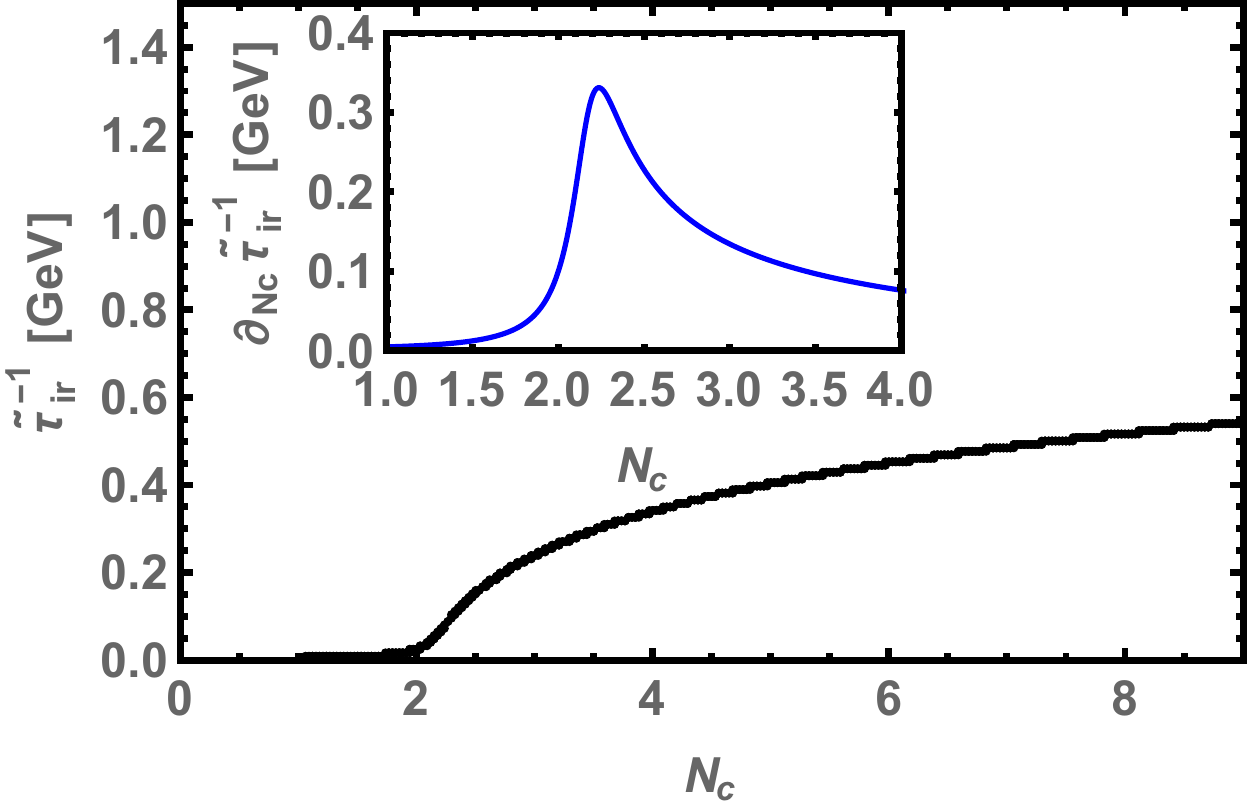}
\caption{}\label{fig:b}
\end{subfigure}
\caption{(\subref{fig:a}) Behavior of the inverse of confining length scale $\tilde{\tau}^{-1}_{ir}$ and its flavor-gradient $\partial_{N_f}\tilde{\tau}^{-1}_{ir}$ as a function of number of flavors $N_f$  for fixed $N_c=3$.  (\subref{fig:b}) The behavior  $\tilde{\tau}^{-1}_{ir}$ and its color-gradient ~$\partial_{N_c}\tilde{\tau}^{-1}_{ir}$  as a function of the number of colors $N_c$ and  for fixed $N_f=2$.}
\label{fig2}
\end{figure}

\begin{figure}
\begin{subfigure}[b]{7cm}
\centering\includegraphics[width=8cm]{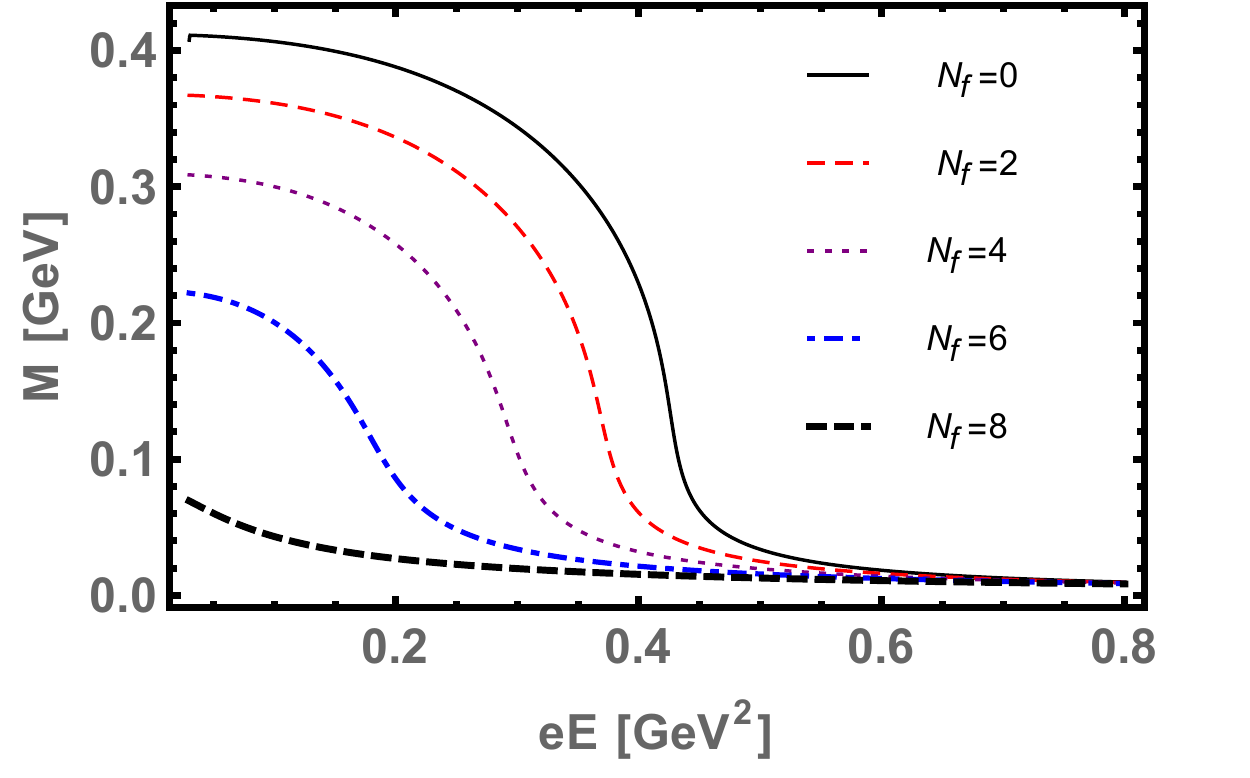}
\caption{}\label{fig:a}
\end{subfigure}
\hfill
\begin{subfigure}[b]{7cm}
\centering\includegraphics[width=8cm]{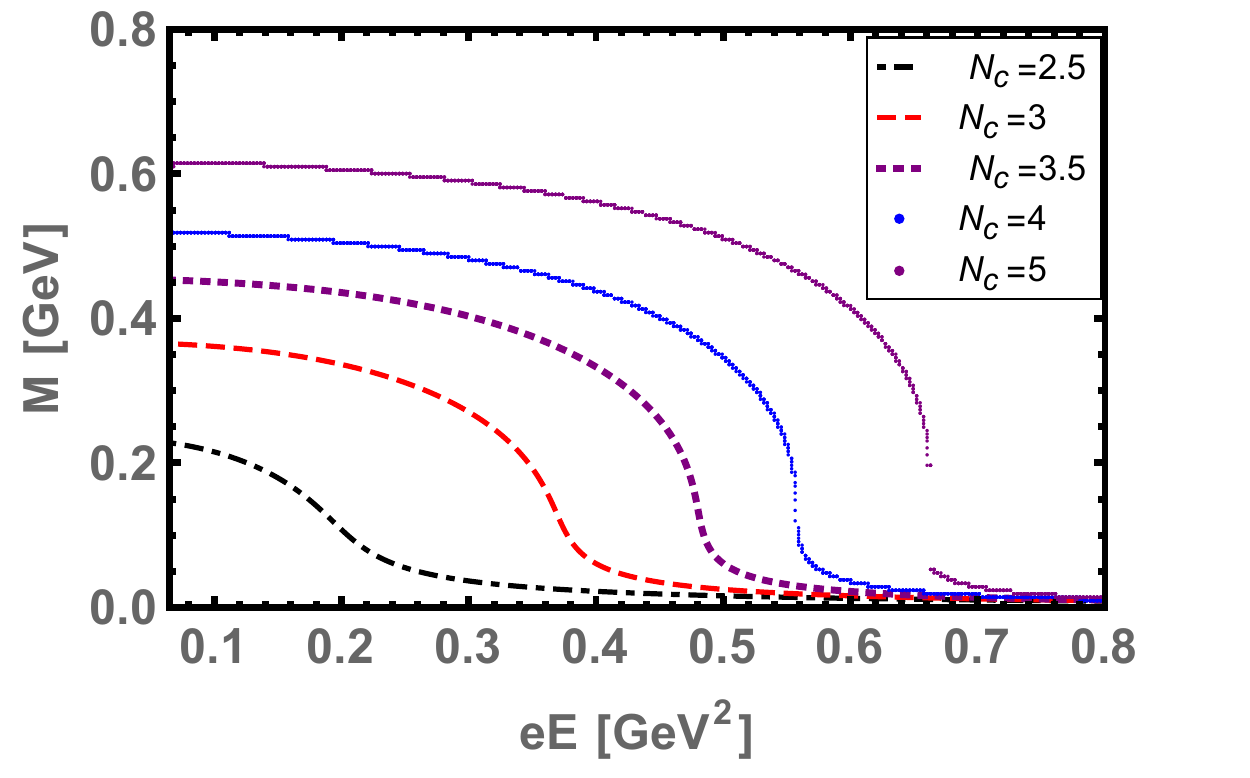}
\caption{}\label{fig:b}
\end{subfigure}
\caption{(\subref{fig: a}) The dynamical quark mass as a function of electric field strength $eE$,  for various  $N_f$ and for fixed $N_c=3$. The plateau of the dynamical mass suppresses as we increase $N_f$. (\subref{fig:b}) shows the behavior of the dynamical quark mass as a function of electric field strength $eE$ for various  $N_c$ and for fixed $N_f=2$. The increasing $N_c$ enhances the plateau of the dynamical mass as a function of $eE$. For $N_c\geq4$, the cross-over  phase transition changes to the first order. 
}
\label{fig3}
\end{figure}

In this section, we present the numerical results of the gap equation at zero electric field and in the presence of electric field for higher $N_c $ and $N_f$.  We use the following set of flavor-dependent contact interaction  model parameters~\cite{Ahmad:2020jzn} i.e., $\tau_{{ir}} = (0.24~\mathrm{GeV})^{-1}$, $\tau_{uv} = (0.905~\mathrm{GeV})^{-1}$,
and  bare quark mass $m_u=m_d=0.007$~GeV. For fixed $N_c=3$ and $N_f=2$, we have obtained the dynamical mass $M_{u,d}=0.367$~GeV and the
condensate $-\langle\bar{q}q\rangle^{1/3}_{u,d}=
0.243$~GeV$^3$. It should be noted that these parameters were fitted to reproduce the light mesons properties~\cite{GutierrezGuerrero:2010md}. \\ Next, we solve the gap equation Eq.(\ref{CI11}) for various  $N_f$ and at  fix $N_c=3$ as shown in the  Fig.~\ref{fig1}. The dynamical mass monotonically decreases as we increase the $N_f$ as depicted in Fig.~\ref{fig1}(a). The plot inside the Fig.~\ref{fig1}(a), represents the flavor-gradient of the dynamical mass~$\partial_{N_f}M_f$, whose peak is at $N^{c}_{f}\approx8$, which is a critical number of flavors where near and above the dynamical chiral symmetry restored. In Fig.~\ref{fig1}(b), we show the dynamical mass as a function of the number of colors $N_c$ for fix  $N_f=2$. This plot represents that the dynamical chiral symmetry broken near or above some critical value of $N^{c}_c\approx 2.2$. The $N^{c}_c\approx 2.2$ is obtained from the peak of the color-gradient of the mass function~$\partial_{N_c}M_f$. These findings are consistent with results obtained in~\cite{Ahmad:2020jzn, Ahmad:2022hbu}. 
The inverse of the confining length scale $\tilde{\tau}^{-1}_{ir}$  and its flavor-gradient plotted for various $N_f$ and for fixed $N_c=3$  are plotted in Fig.~\ref{fig2}(a).  The confinement can be triggered from flavor-gradient~$\partial_{N_f}\tilde{\tau}^{-1}_{ir}$, whose peak is at $N^{c}_{f}\approx8$, and  is approximated as a critical number of flavors above which the quark become unconfined.  In Fig.~\ref{fig2}(b), we show the inverse  confining length scale $\tilde{\tau}^{-1}_{ir}$ as a  function of the various number of colors $N_c$ for fix  $N_f=2$. This plot represents that the confinement occurs near and above at some critical value of $N^{c}_c\approx 2.2$. The critical $N^{c}_{c}\approx 2.2$ for the confinement is obtained from the peak of the color gradient~$\partial_{N_c}\tilde{\tau}^{-1}_{ir}$.
Next, we solve the gap equation Eq.~(\ref{em4}), in the presence of electric field  $eE$  and plotted the dynamical mass as a function of $eE$,  for the various number of flavors $N_f$ and for fix $N_c=3$ as depicted in Fig.~\ref{fig3}(a). The dynamical mass decreases as we increase the electric field $eE$ as  we expected. Upon increasing the $N_f$, the plateau of dynamical quark mass as a function of $eE$ suppresses for larger values of $N_f$. There is no dynamical mass generation above $N^{c}_f=8$, this is because both electric field  $eE$ and  $N_f$ restore the dynamical chiral symmetry and quarks becomes unconfined. \\ In Fig.~\ref{fig3}(b), we plotted the dynamical quark mass as a function of electric field strength $eE$ for various  $N_c$ and for fixed $N_f=2$. The increasing $N_c$ enhances the plateau of the dynamical mass as a function of $eE$. For $N_c\geq4$, the mass plots show the discontinuities in the dynamical symmetry restoration region where the nature of smooth cross-over  phase transition changes to the first order.  This may be due to the strong competition that occurs between $eE$ and $N_c$ (i.e. strong electric field tends to restore the dynamical chiral symmetry whereas larger $N_c$ enhance the dynamical chiral symmetry breaking. 

\begin{figure}
\begin{subfigure}[b]{7cm}
\centering\includegraphics[width=8cm]{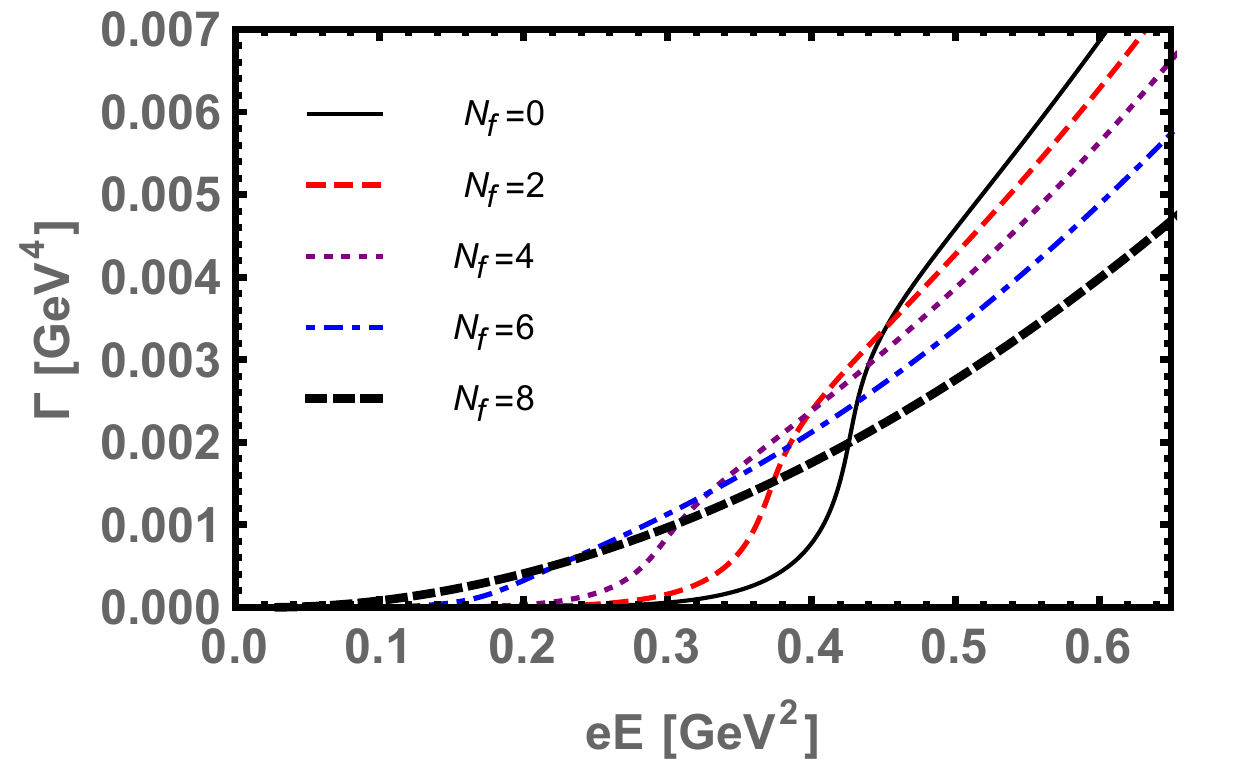}
\caption{}\label{fig:a}
\end{subfigure}
\hfill
\begin{subfigure}[b]{7cm}
\centering\includegraphics[width=8cm]{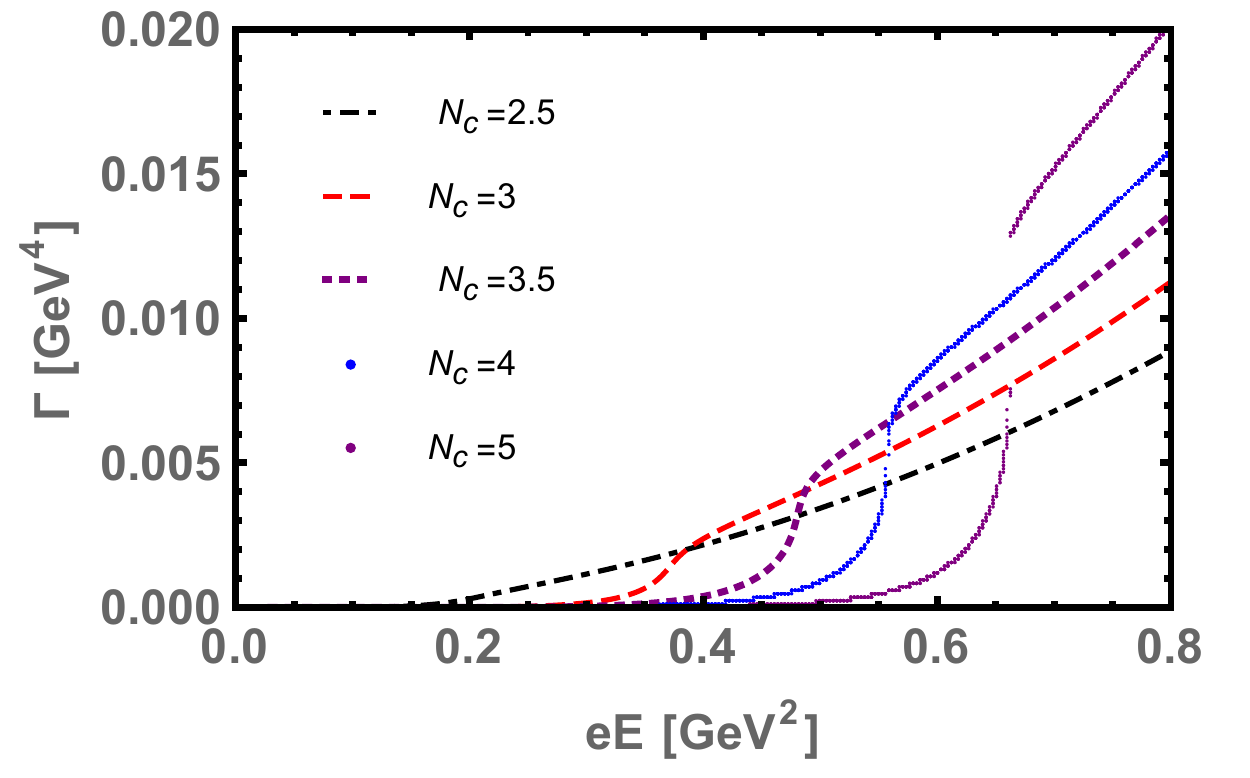}
\caption{}\label{fig:b}
\end{subfigure}
\caption{(\subref{fig:a}) The Schwinger pair production rate $\Gamma$  as a function of electric field strength $eE$,  for various  $N_f$ and for fixed $N_c=3$. Upon increasing $N_f$, the pair production rate grows faster even at small values of $eE$.   (\subref{fig:b}) The behavior of production rate $\Gamma$ as a function of  $eE$ for various  $N_c$ at fixed $N_f=2$. Upon increasing the number of colors $N_c$  the production rate grows slowly and  higher values of  $eE$ needed for quick and stable pair production rate.}
  \label{fig4}
\end{figure}

\begin{figure}
\begin{subfigure}[b]{7cm}
\centering\includegraphics[width=8cm]{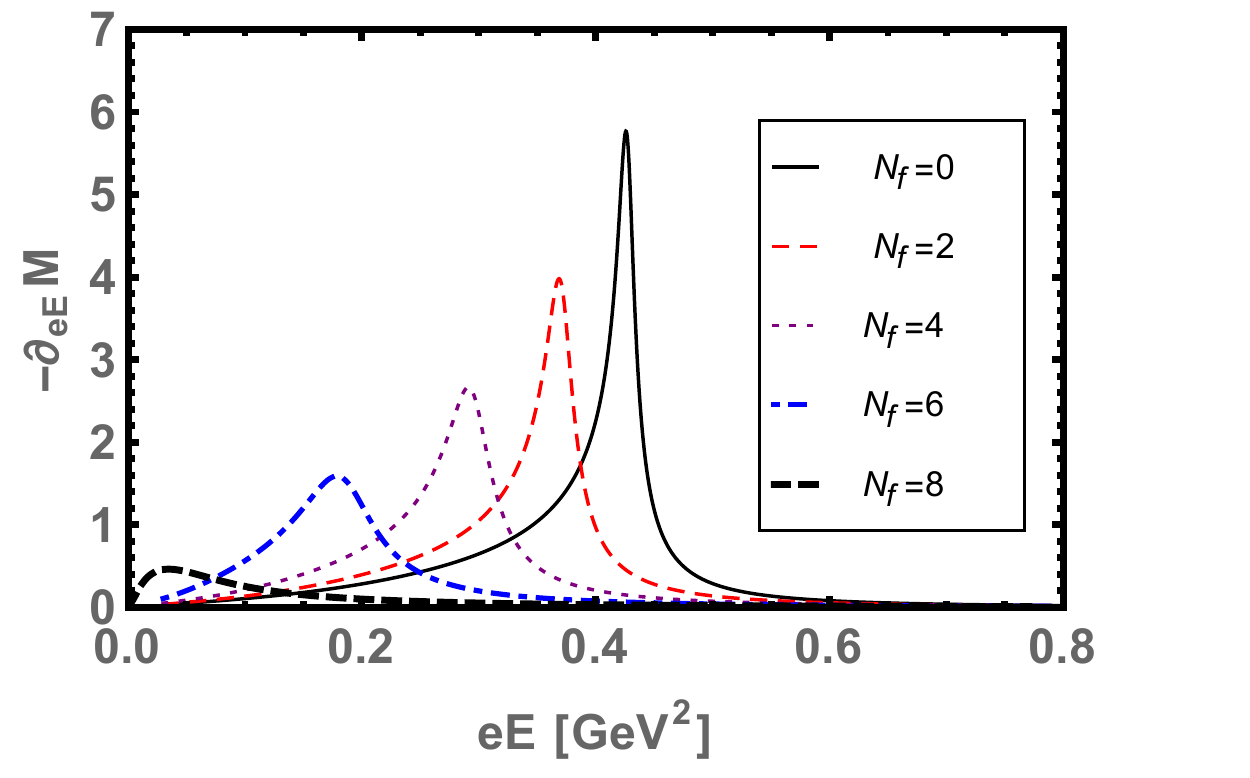}
\caption{}\label{fig:a}
\end{subfigure}
\hfill
\begin{subfigure}[b]{7cm}
\centering\includegraphics[width=8cm]{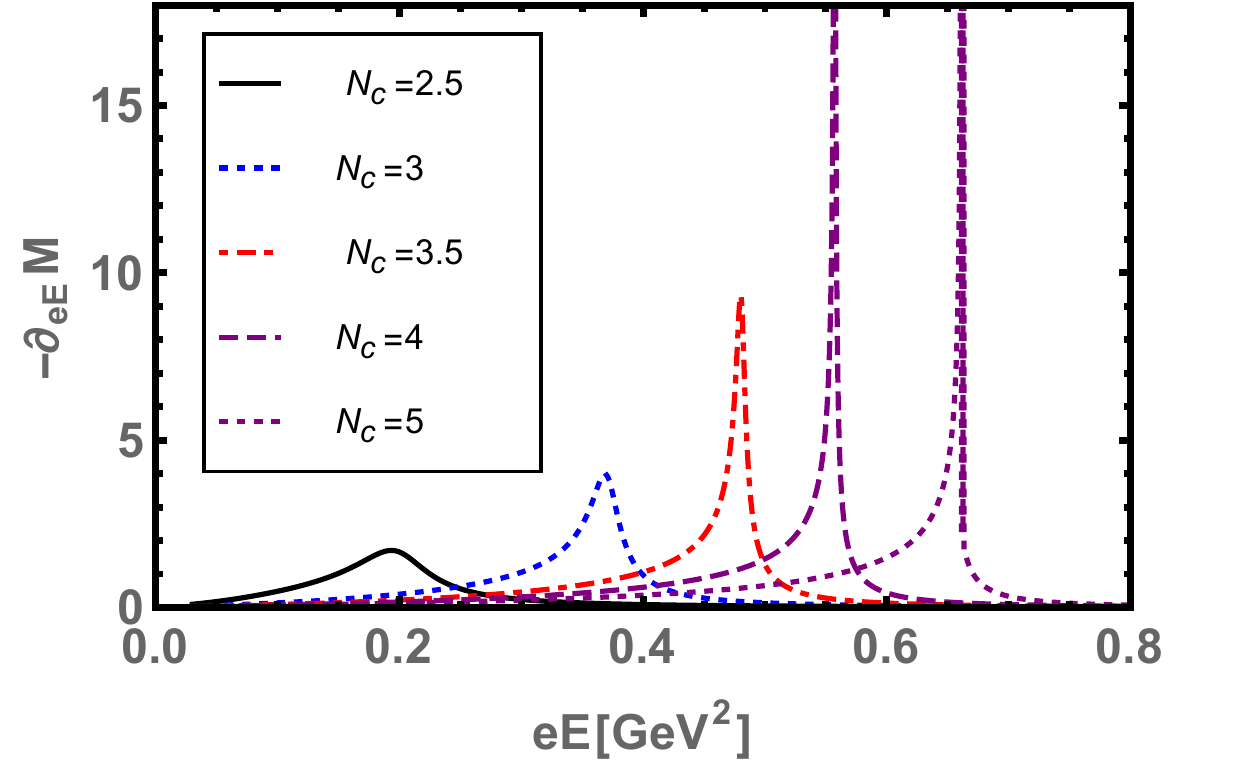}
\caption{}\label{fig:b}
\end{subfigure}
\caption{(\subref{fig:a}) The electric-gradient of the dynamical mass $\partial_{eE}M$ as a function  $N_f$  for fixed $N_c=3$. The peaks shifted towards the lower values of electric field $eE$ upon the increasing $N_f$. (\subref{fig:b}) The electric-gradient  $\partial_{N_c}M$ as a function of  $N_c$  for fixed $N_f=2$.  The peak of the  $\partial_{N_c}M$, shifts toward higher values of $eE$ as $N_c$ increases. For $N_c\geq4$, the peaks in $\partial_{N_c}M$ diverges and dicontinous.}
  \label{fig5}
\end{figure}

\begin{figure}
\begin{subfigure}[b]{7cm}
\centering\includegraphics[width=8cm]{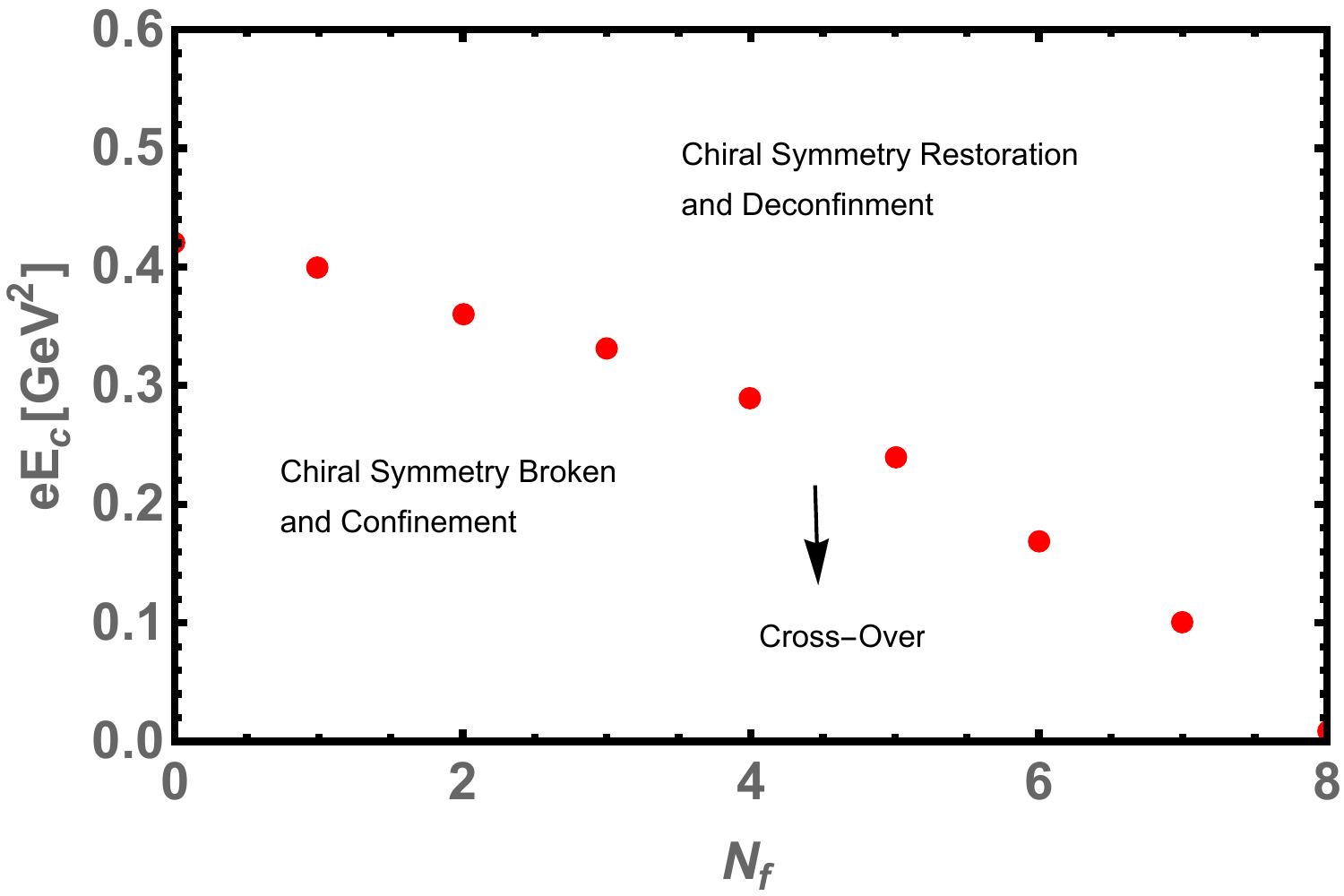}
\caption{}\label{fig:a}
\end{subfigure}
\hfill
\begin{subfigure}[b]{7cm}
\centering\includegraphics[width=8cm]{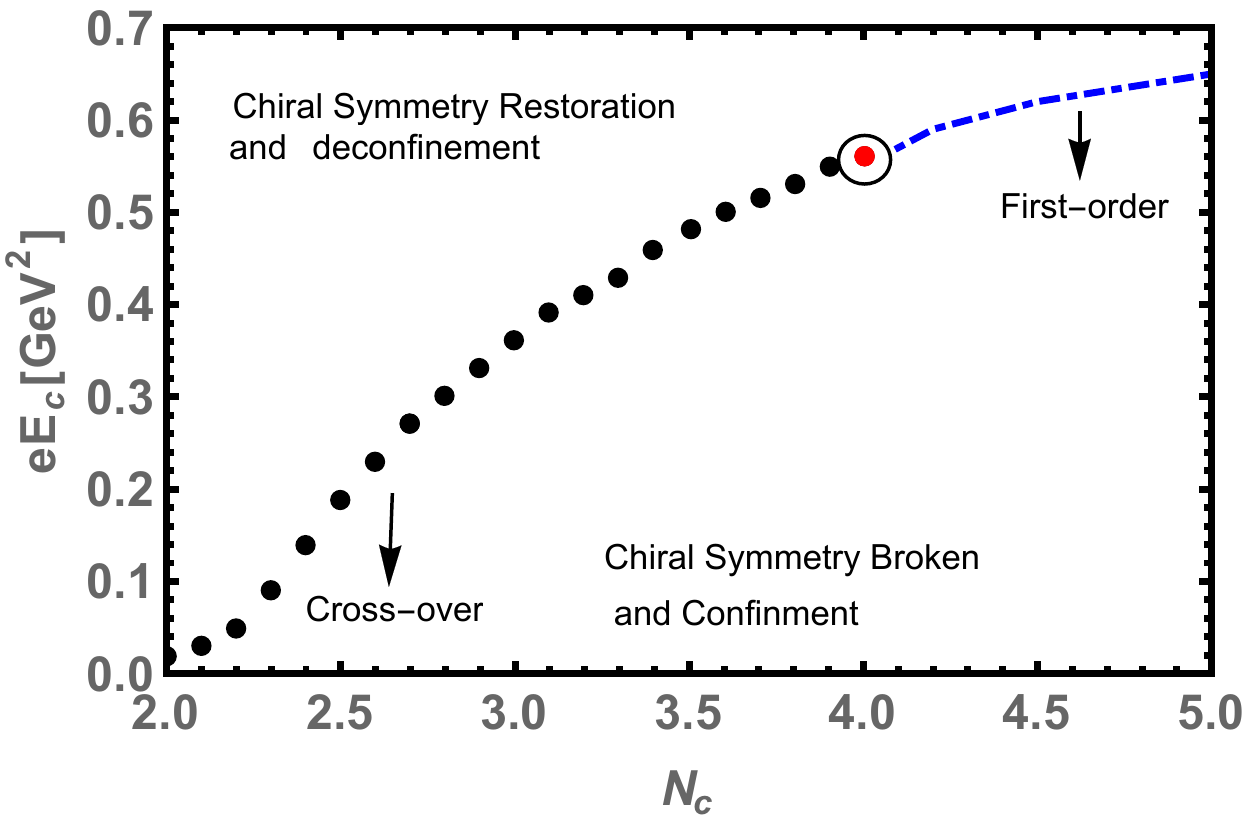}
\caption{}\label{fig:b}
\end{subfigure}
\caption{(\subref{fig:a}) The phase diagram  $N_f-eE_c$ plane for the dynamical chiral symmetry breaking/restoration or confinement-deconfinement transition. The pseudo-critical $eE_c$ decreases upon increasing the number of light quark flavors $N_f$.  (\subref{fig:b}) The phase diagram in $N_c-eE_c$ plane; here the critical $eE$ increase upon increasing the number of color $N_c$. The transition is of smooth cross-over until the critical endpoint $(N^{c}_{c,p}, ,eE_{c,p})$ where it suddenly changes to the first-order.}
\label{fig6}
\end{figure}

The Schwinger pair production rate ``$\Gamma$'' Eq.~(\ref{em9}),  as a function of electric field strength $eE$,  for various  $N_f$ and for fix $N_c=3$ is shown in the Fig.~\ref{fig4}(a). This figure clearly demonstrate that 
after some critical value  $eE_{c}$, where near at and above the dynamical symmetry  restored and the deconfinement occurred, the pair production rate ``$\Gamma$'' grows
faster due to the weakening of the chiral condensates. In this situation, the QCD vacuum becomes more unstable
and the pair of quark-antiquark becomes more likely to
produces. Upon increasing the number of flavors $N_f$, the pair production rate grows faster and shifted towards the lower values of electric field  $eE$. This is because, both the $eE$ and $N_f$ restore the  dynamical chiral symmetry. Although, there is an unstable slow enhancement of pair production rate for small $N_f$ but become stable and faster for higher values of $N_f$   In Fig.~\ref{fig4}(b), we plotted the quark-antiquark pair production  rate``$\Gamma$''  as a function of $eE$ for various  $N_c$  but at this time we fix the number of flavors $N_f=2$. Upon increasing the number of colors $N_c$,  the production rate grows slowly and higher  values of $eE$  are needed for a stable and faster production rate. This is because, $N_c$ enhances the dynamical chiral symmetry breaking. For $N_c\geq4$, the discontinuity occurs in the production rate near and above the chiral symmetry restoration and deconfinement region. This may be due to the strong competition between $N_c$ and $eE$, i.e., on one hand, the strong electric field $eE$  tends to restore the dynamical chiral symmetry whereas the $N_c$ tends to break it.
We then obtained the pseudo-critical electric field $eE_c$ for the chiral symmetry breaking -restoration from the electric field-gradient~$\partial_{eE}M_f$ as a function of $eE$ for various $N_f$ and at fixed $N_c=3$, as depicted in Fig.~\ref{fig5}(a).  The peaks of the ~$\partial_{eE}M_f$ shift towards lower values of the electric field $eE$.  
In Fig.~\ref{fig5}(b), we plotted the electric field gradient ~$\partial_{eE}M_f$ as a function of $eE$ for various $N_c$ and at fixed $N_f=2$. Upon increasing $N_c$, the peaks shift towards  higher values of $eE$. 
We then obtained the pseudo-critical electric field $eE_c$ for the chiral symmetry restoration/deconfinement different $N_f$, and draw a  phase diagram in $N_{f}-eE$ plane a shown in the Fig.~\ref{fig6}(a). The pseudo-critical electric field $eE_c$ decreases as we increase $N_f$, the nature of the chiral phase transition is of smooth cross-over. In Fig.~\ref{fig6}(b), we sketch the phase diagram in $N_{c}-eE_c$ plane, the critical $eE_c$ grows upon increasing the number of colors $N_c$. The transition is of smooth cross-over until the critical endpoint $(N_{c,p}=4, eE_{c,p}=0.54 GeV^2)$  where it suddenly changes to the first-order. The phase diagram also shows that the quark-antiquark pair production grows quickly after pseudo-critical electric field $eE_c$ and how it varies with the flavors and colors. Our finding for $N_c=3$ and $N_f=2$, agrees well with the growth of the quark-antiquark production rate studied through another effective model of low energy QCD~\cite{Cao:2015xja, Tavares:2018poq}. 
In the next section, we discuss the summery and future perspectives of the this work.
\section{Summery and Perspectives}
In this work, we have studied the  impact of 
pure electric 
field on the color-flavor chiral phase transitions and investigated the   Schwinger quark-antiquark pair production rate $\Gamma$ for the higher number of colors $N_c$ and number flavors $N_f$. For this purpose, we
implemented the Schwinger-Dyson formulation of QCD,
with a gap equation kernel comprising a symmetry-preserving
vector-vector flavor-dependent contact interaction model of quarks
in a rainbow-ladder truncation. Subsequently, we adopted
the well-known Schwinger proper-time regularization
procedure. The outcome of this study is presented as
follows:\\
First, we reproduced the results of dynamical chiral symmetry restoration for large $N_f$  with fixed $N_f=3$, where we evaluated the critical number of flavors $N^{c}_{f}\approx 8$. Further, we explore the dynamical symmetry breaking for a higher number of colors $N_c$ but kept $N_f=2$ and found the critical number of colors $N^{c}_c\approx2.2$.
Second, we consider the influence of the pure electric field $eE$, where we have explored the dynamical chiral symmetry restoration and deconfinement for various numbers of flavors $N_f$ and colors $N_c$.  The plateau of the mass as a function of the electric field is noted to suppress upon increasing the number of flavors. Whereas upon increasing the number of colors, the dynamical mass as a function of $eE$  enhances and at some $N_c\geq4$, we found the discontinuity in the chiral symmetry restoration region.
Next, We obtained the  Schwinger pair production (quark-antiquark) rate $\Gamma$ as a function of the pure electric field $eE$ for various $N_f$ and $N_c$. For fixed $N_c=3$, and upon varying the  $N_f$, we found that quark-antiquark production rate $\Gamma$  grows quickly
when we cross a pseudo-critical electric field  $eE_c$. As a result, the 
pair production rate 
tends to initiates at lower values of  $eE$ for larger values of $N_f$. Hence, the pseudo-critical electric field $eE_c$ reduced in magnitude upon increasing the number of flavors $N_f$. This is because both $N_f$ and $eE$ restored the dynamical chiral symmetry. While for fixed $N_f=2$ and upon increasing $N_c$,  the Schwinger pair production
tends to initiate at larger values of $eE$ for higher values of $N_c$.  It is interesting to note that for $N\geq4$, the discontinuity occurred in the production rate in the region where the chiral symmetry is restored. This may be due to the strong competition that occurred between $eE$ and $N_c$. Thus, the pseudo-critical $eE_c$  enhances as the number of colors $N_c$ increases.  The transition is of smooth cross-over until the critical endpoint $(N_{c,p}=4, eE_{c,p}=0.54 GeV^2)$,  where it suddenly changes to the first order.  Our finding for $N_c=3$ and $N_f=2$ agrees well with the behaviour of the pair production rate studied through other effective models of low energy QCD. 
Qualitatively and quantitatively, the predictions of the
presented flavor-dependent contact interaction model for fixed $N_=3$ and $N_f=2$  agree well with results obtained from
other effective QCD models. Soon, we plan to extend this
work to study the Schwinger pair production rate in hot and dense matter QCD in the presence of background fields. We are also interested in extending this work
to study the properties of light hadrons in the
background  fields etc.
\section{Acknowledgments}
We acknowledge A. Bashir and A. Raya for  their guidance and  valuable suggestions  in the process of completion of this work. We also thank to the colleagues of the Institute of  Physics, Gomal University (Pakistan).
\section*{References}
\bibliographystyle{iopart-num}
\bibliography{aftabahmad}

\providecommand{\newblock}{}
\begin{thebibliography}{10}
\expandafter\ifx\csname url\endcsname\relax
  \def\url#1{{\tt #1}}\fi
\expandafter\ifx\csname urlprefix\endcsname\relax\def\urlprefix{URL }\fi
\providecommand{\eprint}[2][]{\url{#2}}

\bibitem{Sauter:1931zz}
Sauter F 1931 {\em Z. Phys.\/} {\bf 69} 742--764

\bibitem{Heisenberg:1936nmg}
Heisenberg W and Euler H 1936 {\em Z. Phys.\/} {\bf 98} 714--732
  (\textit{Preprint} \eprint{physics/0605038})

\bibitem{Schwinger:1951nm}
Schwinger J~S 1951 {\em Phys. Rev.\/} {\bf 82} 664--679

\bibitem{Yildiz:1979vv}
Yildiz A and Cox P~H 1980 {\em Phys. Rev. D\/} {\bf 21} 1095

\bibitem{Cox:1985bu}
Cox P~H and Yildiz A 1985 {\em Phys. Rev. D\/} {\bf 32} 819--820

\bibitem{Suganuma:1990nn}
Suganuma H and Tatsumi T 1991 {\em Annals Phys.\/} {\bf 208} 470--508

\bibitem{Suganuma:1991ha}
Suganuma H and Tatsumi T 1993 {\em Prog. Theor. Phys.\/} {\bf 90} 379--404

\bibitem{Tanji:2008ku}
Tanji N 2009 {\em Annals Phys.\/} {\bf 324} 1691--1736 (\textit{Preprint}
  \eprint{0810.4429})

\bibitem{Klevansky:1988yw}
Klevansky S~P and Lemmer R~H 1988 {\em Phys. Rev.\/} {\bf D38} 3559--3565

\bibitem{Klimenko:1992ch}
Klimenko K~G 1992 {\em Theor. Math. Phys.\/} {\bf 90} 1--6 [Teor. Mat.
  Fiz.90,3(1992)]

\bibitem{Klevansky:1992qe}
Klevansky S~P 1992 {\em Rev. Mod. Phys.\/} {\bf 64} 649--708

\bibitem{Babansky:1997zh}
Babansky A~{\relax Yu}, Gorbar E~V and Shchepanyuk G~V 1998 {\em Phys. Lett.\/}
  {\bf B419} 272--278 (\textit{Preprint} \eprint{hep-th/9705218})

\bibitem{Cao:2015dya}
Cao G and Huang X~G 2016 {\em Phys. Rev. D\/} {\bf 93} 016007
  (\textit{Preprint} \eprint{1510.05125})

\bibitem{Tavares:2018poq}
Tavares W~R and Avancini S~S 2018 {\em Phys. Rev. D\/} {\bf 97} 094001
  (\textit{Preprint} \eprint{1801.10566})

\bibitem{Wang:2017pje}
Wang L and Cao G 2018 {\em Phys. Rev.\/} {\bf D97} 034014 (\textit{Preprint}
  \eprint{1712.09780})

\bibitem{Ruggieri:2016lrn}
Ruggieri M and Peng G~X 2016 {\em Phys. Rev.\/} {\bf D93} 094021
  (\textit{Preprint} \eprint{1602.08994})

\bibitem{Tavares:2019mvq}
Tavares W~R, Farias R~L~S and Avancini S~S 2020 {\em Phys. Rev. D\/} {\bf 101}
  016017 (\textit{Preprint} \eprint{1912.00305})

\bibitem{Ahmad:2020ifp}
Ahmad A 2021 {\em Chin. Phys. C\/} {\bf 45} 073109 (\textit{Preprint}
  \eprint{2009.09482})

\bibitem{Bzdak:2011yy}
Bzdak A and Skokov V 2012 {\em Phys. Lett.\/} {\bf B710} 171--174
  (\textit{Preprint} \eprint{1111.1949})

\bibitem{Deng:2012pc}
Deng W~T and Huang X~G 2012 {\em Phys. Rev. C\/} {\bf 85} 044907
  (\textit{Preprint} \eprint{1201.5108})

\bibitem{Bloczynski:2012en}
Bloczynski J, Huang X~G, Zhang X and Liao J 2013 {\em Phys. Lett.\/} {\bf B718}
  1529--1535 (\textit{Preprint} \eprint{1209.6594})

\bibitem{Bloczynski:2013mca}
Bloczynski J, Huang X~G, Zhang X and Liao J 2015 {\em Nucl. Phys.\/} {\bf A939}
  85--100 (\textit{Preprint} \eprint{1311.5451})

\bibitem{Hirono:2012rt}
Hirono Y, Hongo M and Hirano T 2014 {\em Phys. Rev. C\/} {\bf 90} 021903
  (\textit{Preprint} \eprint{1211.1114})

\bibitem{Voronyuk:2014rna}
Voronyuk V, Toneev V~D, Voloshin S~A and Cassing W 2014 {\em Phys. Rev. C\/}
  {\bf 90} 064903 (\textit{Preprint} \eprint{1410.1402})

\bibitem{Deng:2014uja}
Deng W~T and Huang X~G 2015 {\em Phys. Lett. B\/} {\bf 742} 296--302
  (\textit{Preprint} \eprint{1411.2733})

\bibitem{Huang:2013iia}
Huang X~G and Liao J 2013 {\em Phys. Rev. Lett.\/} {\bf 110} 232302
  (\textit{Preprint} \eprint{1303.7192})

\bibitem{Jiang:2014ura}
Jiang Y, Huang X~G and Liao J 2015 {\em Phys. Rev.\/} {\bf D91} 045001
  (\textit{Preprint} \eprint{1409.6395})

\bibitem{Karpenko:2016jyx}
Karpenko I and Becattini F 2017 {\em Eur. Phys. J.\/} {\bf C77} 213
  (\textit{Preprint} \eprint{1610.04717})

\bibitem{Xia:2018tes}
Xia X~L, Li H, Tang Z~B and Wang Q 2018 {\em Phys. Rev.\/} {\bf C98} 024905
  (\textit{Preprint} \eprint{1803.00867})

\bibitem{Wei:2018zfb}
Wei D~X, Deng W~T and Huang X~G 2019 {\em Phys. Rev.\/} {\bf C99} 014905
  (\textit{Preprint} \eprint{1810.00151})

\bibitem{Cao:2015xja}
Cao G and Zhuang P 2015 {\em Phys. Rev. D\/} {\bf 92} 105030 (\textit{Preprint}
  \eprint{1505.05307})

\bibitem{LSD:2014nmn}
Appelquist T {\em et~al.\/} (LSD) 2014 {\em Phys. Rev. D\/} {\bf 90} 114502
  (\textit{Preprint} \eprint{1405.4752})

\bibitem{Hayakawa:2010yn}
Hayakawa M, Ishikawa K~I, Osaki Y, Takeda S, Uno S and Yamada N 2011 {\em Phys.
  Rev. D\/} {\bf 83} 074509 (\textit{Preprint} \eprint{1011.2577})

\bibitem{Cheng:2013eu}
Cheng A, Hasenfratz A, Petropoulos G and Schaich D 2013 {\em JHEP\/} {\bf 07}
  061 (\textit{Preprint} \eprint{1301.1355})

\bibitem{Hasenfratz:2016dou}
Hasenfratz A and Schaich D 2018 {\em JHEP\/} {\bf 02} 132 (\textit{Preprint}
  \eprint{1610.10004})

\bibitem{LatticeStrongDynamics:2018hun}
Appelquist T {\em et~al.\/} (Lattice Strong Dynamics) 2019 {\em Phys. Rev. D\/}
  {\bf 99} 014509 (\textit{Preprint} \eprint{1807.08411})

\bibitem{bashir2013qcd}
Bashir A, Raya A and Rodriguez-Quintero J 2013 {\em Physical Review D\/} {\bf
  88} 054003

\bibitem{Appelquist:1999hr}
Appelquist T, Cohen A~G and Schmaltz M 1999 {\em Phys. Rev. D\/} {\bf 60}
  045003 (\textit{Preprint} \eprint{hep-th/9901109})

\bibitem{Hopfer:2014zna}
Hopfer M, Fischer C~S and Alkofer R 2014 {\em JHEP\/} {\bf 11} 035
  (\textit{Preprint} \eprint{1405.7031})

\bibitem{Doff:2016jzk}
Doff A and Natale A~A 2016 {\em Phys. Rev. D\/} {\bf 94} 076005
  (\textit{Preprint} \eprint{1610.02564})

\bibitem{Binosi:2016xxu}
Binosi D, Roberts C~D and Rodriguez-Quintero J 2017 {\em Phys. Rev. D\/} {\bf
  95} 114009 (\textit{Preprint} \eprint{1611.03523})

\bibitem{Ahmad:2020jzn}
Ahmad A, Bashir A, Bedolla M~A and Cobos-Mart\'\i{}nez J~J 2021 {\em J. Phys.
  G\/} {\bf 48} 075002 (\textit{Preprint} \eprint{2008.03847})

\bibitem{Ahmad:2022hbu}
Ahmad A and Murad A 2022 {\em Chin. Phys. C\/} {\bf 46} 083109
  (\textit{Preprint} \eprint{2201.09980})

\bibitem{Ahmad:2016iez}
Ahmad A and Raya A 2016 {\em J. Phys.\/} {\bf G43} 065002 (\textit{Preprint}
  \eprint{1602.06448})

\bibitem{Marquez:2015bca}
Marquez F, Ahmad A, Buballa M and Raya A 2015 {\em Phys. Lett.\/} {\bf B747}
  529--535 (\textit{Preprint} \eprint{1504.06730})

\bibitem{Langfeld:1996rn}
Langfeld K, Kettner C and Reinhardt H 1996 {\em Nucl. Phys. A\/} {\bf 608}
  331--355 (\textit{Preprint} \eprint{hep-ph/9603264})

\bibitem{Cornwall:1981zr}
Cornwall J~M 1982 {\em Phys. Rev. D\/} {\bf 26} 1453

\bibitem{Aguilar:2015bud}
Aguilar A~C, Binosi D and Papavassiliou J 2016 {\em Front. Phys. (Beijing)\/}
  {\bf 11} 111203 (\textit{Preprint} \eprint{1511.08361})

\bibitem{GutierrezGuerrero:2010md}
Gutierrez-Guerrero L~X, Bashir A, Cloet I~C and Roberts C~D 2010 {\em Phys.
  Rev.\/} {\bf C81} 065202 (\textit{Preprint} \eprint{1002.1968})

\bibitem{Kohyama:2016obc}
Kohyama H 2016  (\textit{Preprint} \eprint{1602.09056})

\bibitem{Boucaud:2011ug}
Boucaud P, Leroy J~P, Yaouanc A~L, Micheli J, Pene O and Rodriguez-Quintero J
  2012 {\em Few Body Syst.\/} {\bf 53} 387--436 (\textit{Preprint}
  \eprint{1109.1936})

\bibitem{Ahmad:2015cgh}
Ahmad A, Ayala A, Bashir A, Gutiérrez E and Raya A 2015 {\em J. Phys. Conf.
  Ser.\/} {\bf 651} 012018

\bibitem{Ebert:1996vx}
Ebert D, Feldmann T and Reinhardt H 1996 {\em Phys. Lett.\/} {\bf B388}
  154--160 (\textit{Preprint} \eprint{hep-ph/9608223})

\bibitem{Ward:1950xp}
Ward J~C 1950 {\em Phys. Rev.\/} {\bf 78} 182

\bibitem{Takahashi:1957xn}
Takahashi Y 1957 {\em Nuovo Cim.\/} {\bf 6} 371

\bibitem{Roberts:2007jh}
Roberts C~D, Bhagwat M~S, Holl A and Wright S~V 2007 {\em Eur. Phys. J. ST\/}
  {\bf 140} 53--116 (\textit{Preprint} \eprint{0802.0217})

\bibitem{Cao:2021rwx}
Cao G 2021 {\em Eur. Phys. J. A\/} {\bf 57} 264 (\textit{Preprint}
  \eprint{2103.00456})

\end{thebibliography}

\end{document}